# CreateAI

Insights from an NSF Workshop on K–12 Students, Teachers, and Families as Designers of Artificial Intelligence and Machine Learning Applications

**Yasmin B. Kafai**, University of Pennsylvania
**José Ramón Lizárraga**, University of California, Berkeley
**R. Benjamin Shapiro**, University of Washington

With
**Marina Bers**, Boston College
**Kathi Fisler**, Brown University

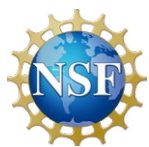

**CITE AS**
Kafai, Y.B., Lizárraga, J.R., & Shapiro, R.B. (2025, October). *CreateAI*: Insights from an NSF Workshop on K–12 Students, Teachers, and Families as Designers of Artificial Intelligence and Machine Learning Applications. Philadelphia, PA: University of Pennsylvania. Available at: https://www.createai-workshop.com/workshop-report

**ACKNOWLEDGEMENTS**
This workshop organized by Yasmin Kafai, Marina Bers, Kathi Fisler, José Lizárraga, and R. Benjamin Shapiro was supported by a grant (#2414590) from the National Science Foundation. Any opinions, findings, and conclusions or recommendations expressed in this report are those of the authors and do not necessarily reflect the views of the NSF or the University of Pennsylvania, Boston University, Brown University, University of Washington, and University of California. All images included in this report were provided by workshop participants.

# CreateAI

Insights from an NSF Workshop on K–12 Students, Teachers, and Families as Designers of Artificial Intelligence and Machine Learning Applications

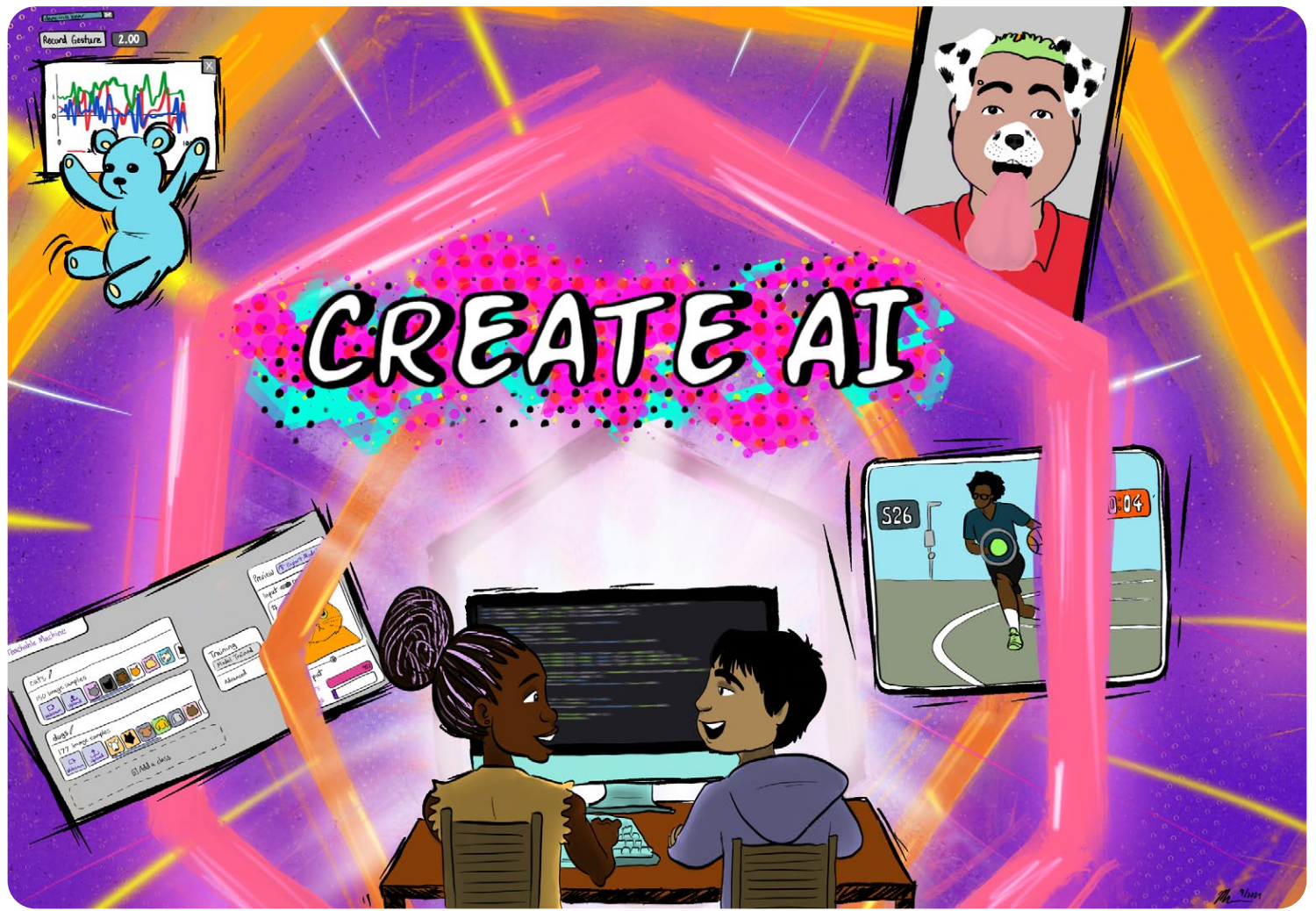

# CONTENTS

# *EXECUTIVE SUMMARY*

K–12 AI initiatives are accelerating worldwide. These initiatives most often promote students, teachers, and administrators as users of artificial intelligence and machine learning (hereafter AI/ML) by presenting personalized learning experiences, implementing adaptive assessments, and fostering administrative efficiency. Yet among these initiatives, few efforts have cast students and teachers as creators who can build, test, and critique AI/ML systems. **What if students and teachers were not just learning to be competent users of AI—but also its creators?** This question is at the heart of CreateAI—a series of NSF-funded meetings that is the basis for this report—in which K–12 educators, researchers, and learning scientists addressed the following questions: (1) What tools, skills, and knowledge will empower students and teachers to build their own AI/ML applications? (2) How can we integrate these approaches into classrooms? and (3) What new possibilities for learning emerge when students and teachers become innovators and creators?

CreateAI addresses these questions by positioning students, teachers, and families as designers who *create AI* and *create with AI*. Across panels and a workshop meeting that took place in Fall 2024 and Spring 2025, participants examined what tools, learning designs, ethical practices, teaching supports, and assessments are needed so learners can build models and applications, understand limitations, and exercise techno-social agency. A key idea that emerged in discussions is that we need to consider two complementary directions for CreateAI: (1) *Creating AI*, in which learners build models and AI/ML applications with an emphasis on data-driven development, training, testing, and ethical reflection and (2) *Creating with AI*, in which learners use AI as a creative medium while maintaining agency through iterative critique and revision. Together, these form a dynamic cycle that cultivates what the workshop called *technosocial change agency*: the capacity to understand how social values and structures shape AI systems, to deconstruct and redesign those systems, and to imagine and work toward more ethical, inclusive technological futures.

The key insights and recommendations that were generated in workshop discussions target tool design, student learning, ethical issues, teaching challenges, and the need for assessment:

**TOOLS**
- Design for tradeoffs among malleability/rigidity, robustness, and composability.
- Design for accessibility from the outset.
- Build progressions that let novices advance without "sharp cliffs."
- Help learners and teachers select appropriate tools.
- Pair any new tools/curricula with robust teacher support.
- Document and disclose AI features, expected use contexts, and limitations.
- Support linguistic/cultural heterogeneity and accessibility.
- Expand support for collaborative modeling and avoid policies that penalize non-use of AI when inappropriate.

**LEARNING**
- Leverage learners' funds of knowledge and add scaffolds for dataset design, model training, and testing.
- Counter mystification by cultivating scientifically grounded mental models of AI systems.
- Engage youth in authentic, empowering ethical inquiry throughout design.
- Position students as co-creators and critics.

**ETHICS**
- Integrate ethics across curricula, and not as an add-on, to discuss data origins, model use, embedded values, and distributed responsibility.
- Use code-free approaches such as algorithm auditing, and create avenues for sharing audit results.
- Attend to representation and bias, supporting cultural agency in design.
- Confront AI's environmental impacts by engaging learners in discussion and consideration of ecological footprints.

**TEACHING**
- Provide professional development, materials, and tools.
- Research age-appropriate pedagogies for diverse contexts.
- Clarify choices among data-driven vs. rule-based and no-code vs. coding approaches.
- Use challenge-centered pedagogies (including deliberately "breaking" systems) to surface limitations.
- Embed ethics and sociotechnical critique in teacher learning.
- Address policy fragmentation so schools know what can be adopted where.
- Align with evolving standards work (e.g., CSTA revisions toward 2026).

**ASSESSMENT**
- Establish baselines of core AI/ML concepts and competencies that learners need to master across grade bands.
- Employ multiple instruments: student self-assessments, curated portfolios, peer audits, and stakeholder-matrix reflections on functionality and ethics.
- Develop assessments that promote students as responsible creators, not just technicians.

# INTRODUCTION

Artificial intelligence and machine learning (AI/ML) are expanding at an exponential pace prompting calls for more robust AI literacy around the globe. National priorities, as outlined in the U.S. government's AI Action Plan, call for cultivating AI talent and broadening participation in AI development—not just use—across all sectors, including education (U.S. Government, 2023). Internationally, we also see significant investment in various countries. For instance, India included funds in their 2025 budget prioritizing up-skilling their workforce and building an infrastructure for building excellence in AI in education (IndiaAI, 2025). Similarly, China has mandated AI education across all grade levels (Reuters, 2025), and the United Kingdom aims to increase AI professionals by the tens of thousands by 2030 (UK Government, 2025). The proposed frameworks vary—some stress key competencies like perception, reasoning, and societal impact (Long & Magerko, 2021; Touretzky et al., 2019; Touretzky & Gardner-McCune, 2022) while others focus on computational empowerment through the construction and deconstruction of technology (Dindler et al., 2020). Regardless of the framework, students, teachers, and families (Druga et al., 2021) need to comprehend how AI/ML is used in various situations and how it is shaping society. All K-12 students and teachers need to be prepared to understand, use, and critique AI/ML applications, an objective also endorsed in national (CSTA & AI4K12, 2025) and global initiatives (UNESCO, 2023).

In reality, however, most AI-in-education efforts position learners as recipients of instruction from AI agents or collaborators with AI agents (Eisenberg et al., 2017; Ouyang & Jiao, 2021) rather than as creators. Much less attention has been given to how learners create, train, test, and research their own AI/ML applications (see also Druga, 2018). Like Touretzky and colleagues (2022), we see the construction of computational artifacts using AI/ML-based applications as a premier opportunity for students to see themselves as "not just programmers, but AI application developers" (p. 10). While K-12 computing education research has recognized the potential for consequential computational action with a variety of curricular activities and programmable tools (Issa et al., 2025; Kafai & Burke, 2014; Kafai & Proctor, 2022; Shapiro & Tissenbaum, 2019), much less is known about what this means in the context of developing applications involving AI/ML. Most of these efforts have been limited to professionals and university students with more advanced technical skills. However, as the OECD AI Literacy Framework (2025) highlights, *designing* AI systems and *creating with AI* are crucial for learners to understand how AI works and its implications in the world.

In this report we outline potential directions and critical issues for bringing students as designers of AI/ML applications into K-12 education. The report is based on a series of panel discussions that started in Fall 2024. We organized three virtual panels—*Learning and Teaching with Creating AI* (October 4, 2024), *Designing Tools for Creating AI* (October 18, 2024), and *Envisioning Learning Futures with Creating AI* (December 14, 2024)—followed by a public poster session on March 2, 2025, an in-person workshop on March 3-4, 2025, at the University of Pennsylvania in Philadelphia, PA, and an invited symposium at the International Society of Learning Sciences 2025 conference. These meetings brought together a group of computing education and learning scientists, literacy scholars, teachers, and AI designers selected to represent a mix of academic disciplines, industry, and professional organizations. Due to the novelty of the research area, a significant number of participants were in early stages of their careers. The core PI team represented various positions, seniority levels, and institutions, which allowed

*Illustration by Mia Shaw*

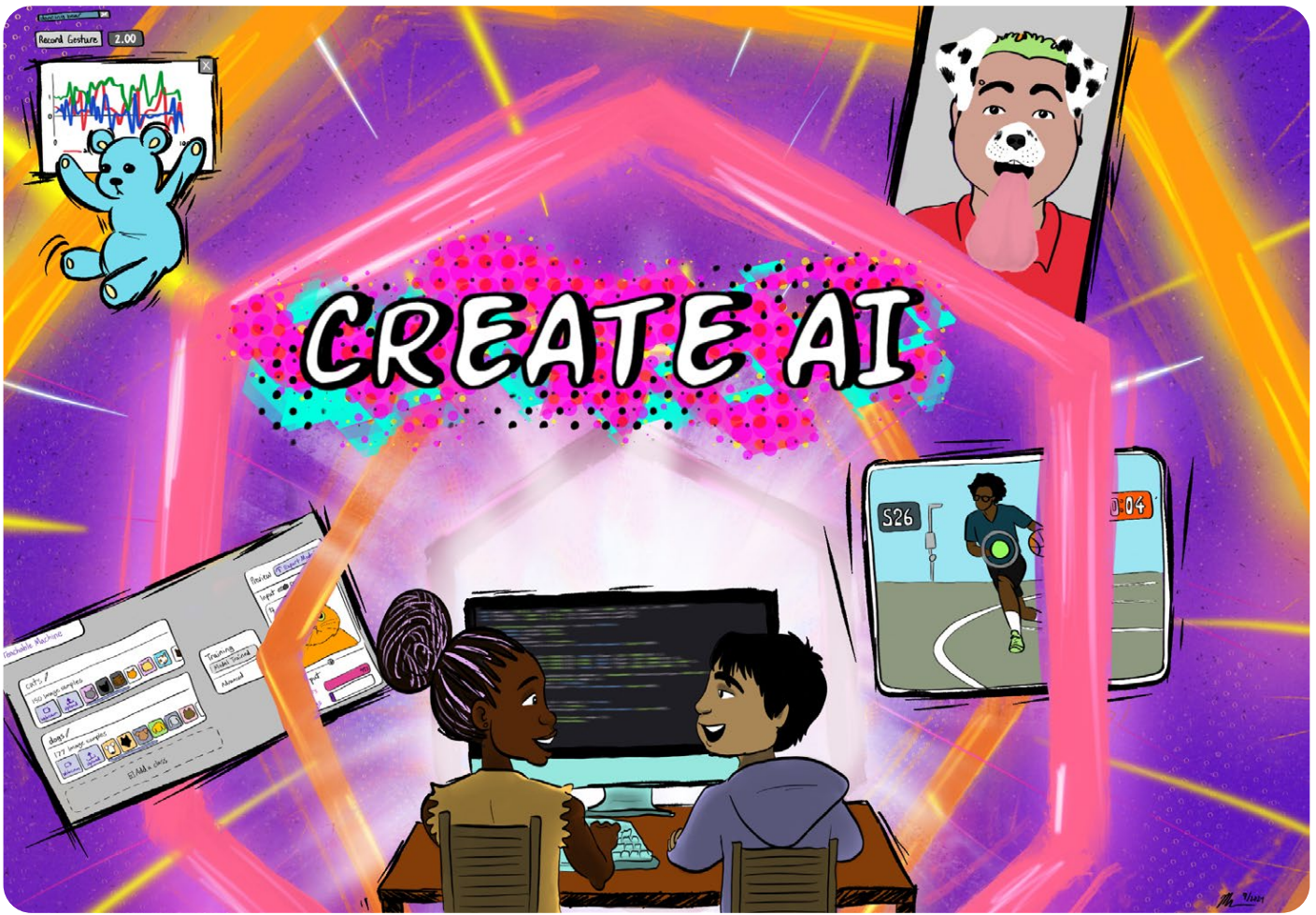

us to effectively leverage our diverse networks for the purposes of recruitment. (For more information on meetings, participants, and agendas, see Appendices A, B, C, and D.)

In workshop discussions, we addressed different dimensions of what learners can gain from designing AI/ML applications, examined how students can develop a better understanding of how AI/ML systems work, engaged critically through specific learning activities, and identified challenges such as institutional constraints. Equally important are considerations of how and when to use AI, and what happens when people elect to not to use AI. Recent research suggests that the use of AI/ML in professional contexts is not always warranted since it does not necessarily result in productivity gains (Becker et al., 2025). Furthermore, numerous studies show a "gender gap" in gen AI use (Deloitte, 2024). This report focuses on a set of intersecting ideas covering the design of tools supporting learners in the process, ethical issues associated with AI/ML, what learners have to understand about AI/ML, and how teachers can support learners and also assess them in this process.

In Section 1, we expand on the different directions of what CreateAI can mean. In Section 2, we provide more detail on how tool design, student learning, critical literacies, teaching, and assessment can be framed for CreateAI. More specifically, in the **TOOLS** section (§2.1) we focus on encouraging student agency in technological innovation, bridging technical complexity with age-appropriate learning by developing adaptive tools that evolve with technological advances while also fostering critical thinking about AI design and capabilities. In the **LEARNING** section (§2.2), we argue for supporting students in developing scientifically grounded understandings of AI, moving beyond mystification to critical engagement and potential system reinvention. In the discussion of **ETHICS** (§2.3), we move beyond technical skills to discuss how learners develop ethical awareness and how we may support them to gain a nuanced understanding of AI's potential and limitations, thereby ensuring inclusive and motivating educational experiences. In the **TEACHING** section (§2.4), we identify critical areas of pedagogy for meaningful, equitable, and forward-thinking AI education approaches that empower students across diverse backgrounds. In **ASSESSMENT** (§2.5), we focus on better understanding how teachers could assess student learning of AI key concepts and practices in creating AI applications. Finally, in Section 3, we review further directions for K-12 education and research.

Throughout the report, we highlight examples of tools and activities that have realized aspects of CreateAI. In the Appendices we provide more detail on the panelists and workshop participants who contributed to the discussions and this report.

# 1. WHAT IS CREATE AI?
## two possible directions

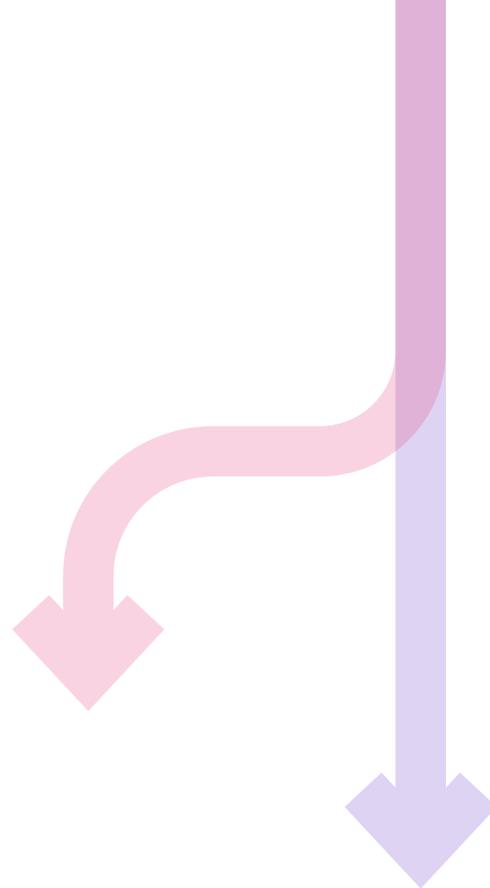

The goal of CreateAI is to empower learners to create, train, evaluate, and research AI/ML applications. Thus, the CreateAI workshops, panels, and symposium brought together researchers, designers, and educators whose work emphasizes agency, creativity, critical thinking, and hands-on design and development over passive consumption of AI technology. Discussions centered on two distinct approaches to CreateAI: *Creating AI* and *Creating with AI*.

The first approach, *Creating AI*, involves having students build AI systems rather than just using AI for learning. As one workshop participant stated, "When we talk about AI education and literacy, it's not that we use AI to teach kids. It's actually that kids build the AI and decide how to train those models, what to do with the trained models, what questions to ask, [and] when and if we should use it." For example, students might design a small generative language model—a "babyGPT"—for writing Marvel scripts (Morales-Navarro et al., 2025; see Figure 1). Or learners can collaboratively train and test image classifiers and build mobile applications that use them (Tseng et al., 2024). A key element in all such projects is having learners use data rather than rules for creating AI/ML applications. In the words of one participant, "We go all data-driven. And there is so much you can do without actually having coding in there… creating media using all kinds of sensors to actually build stuff yourself."

The other, equally important, approach of *Creating with AI* positions AI technologies as new media for learners to express their creativity. As one workshop participant stated: "When I was able to suddenly draw things using this new medium. I'm really bad at drawing, but now I could actually make stuff that is better than what I would do." For example, Scratch Copilot (see Figure 2) can be used as a "medium for creativity" by helping youth realize their programming ideas (Druga & Ko, 2025).

Taken together, both CreateAI approaches can be seen as a dynamic cycle, where *Creating with AI* and *Creating AI* activities are interconnected, with feedback flowing

## Building babyGPTs to Write Screenplays

Morales-Navarro and colleagues (2025) presented a case study of three teenagers (ages 14-15) building a babyGPT screenplay generator. They illustrate how the team developed a model while engaging in data practices relevant to artificial intelligence/machine learning data and addressing ethical issues. Building babyGPTs supported youth to question the "aesthetic legitimacy" of outputs and develop nuanced understandings of the functionality and ethics of generative language models. The teens developed functional understandings related to how the quality and quantity of the training data and the training process influence the quality of the outputs. They considered ethical issues related to the trustworthiness of model outputs, the appropriate use of language models, authorship, and the ownership of creative ideas.

**Figure 1:** *Creating AI:* Building babyGPTs to Write Screenplays (Morales-Navarro et al., 2025)

## Copilots for Creative Coding

Druga and Ko (2025) introduced a prototype AI assistant called Cognimates Copilot to help young learners using Scratch comprehend code, debug programs, and generate new ideas for projects. Their work expands on [cognimates.me](cognimates.me), an open-source platform for AI education. In an exploratory qualitative evaluation with 18 international children (ages 7-12), they found that the AI assistant supported key creative coding processes, particularly aiding ideation and debugging. Crucially, the study highlights how children actively negotiated the use of AI, demonstrating strong agency by adapting or rejecting suggestions to maintain creative control.

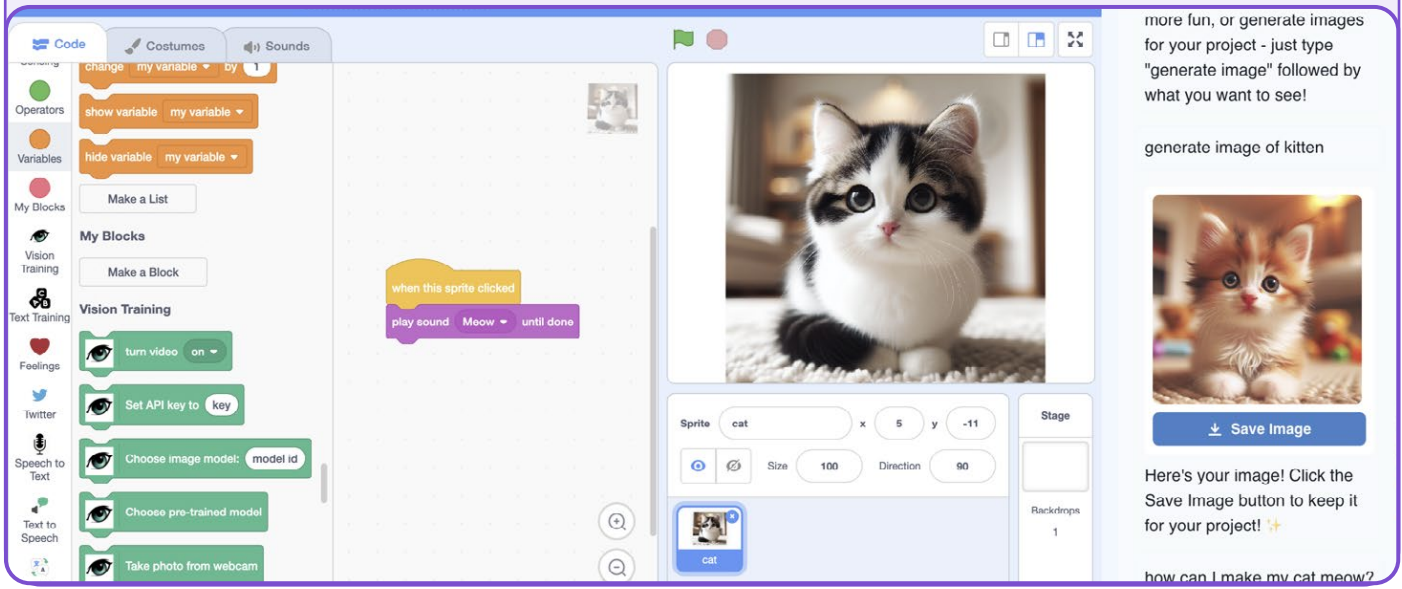

**Figure 2:** *Creating with AI:* Scratch Co-Pilot (Druga & Ko, 2025)

between these processes. One working group captured this sentiment in their proposed definition: "*CreateAI* is an inclusive social setting where AI, along with other resources, supports human interactions and collective reflections, enabling teachers and students to express themselves creatively in ways that would otherwise be challenging."

In both *Creating AI* and *Creating with AI* approaches, learning takes place through iteration and critical engagement, as one workshop participants explained: "[For] the kids creating with generative AI, that back and forth is so important to them having agency, prompting one way and then saying, 'No, this isn't what I want.' …. 'I don't like this output. Let me try and change it.'" Furthermore, there is a need for critique of the system itself: "I'm gonna have agency to create, maybe in collaboration with adults or teachers, to create a new system or add on to a system." The overall goal is to help learners use the creative process to develop *technosocial change agency*, which one participant described as "the ability to understand how society impacts its embedded AI systems, … deconstruct those [systems], and consider … what a more ethical future [could] look like."

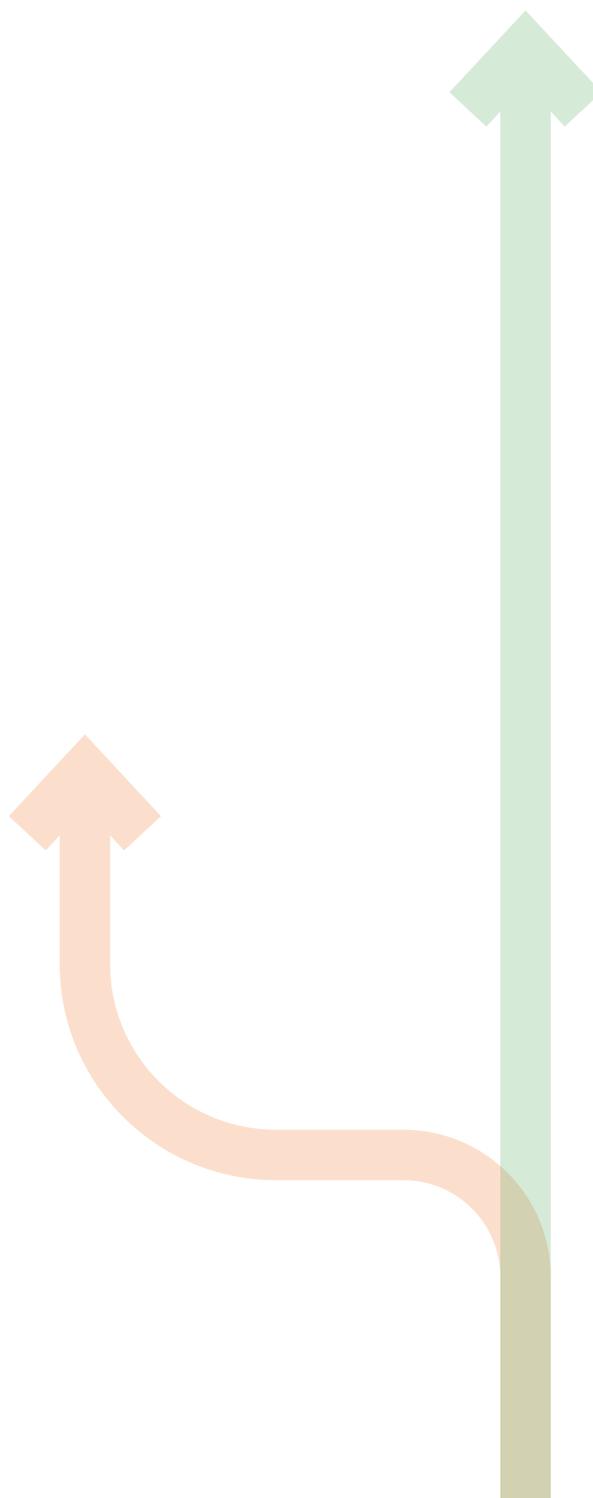

# 2. CREATE AI IN K-12 EDUCATION

# 2.1. TOOLS

Numerous tools for K-12 AI education already exist, some specifically developed for educational purposes and others appropriated by educators and researchers. These tools fall into two general categories. In the first category are extensions to existing computer programming infrastructure that support users in adding ML-based functionality to their projects. Some of these tools cannot be customized by users (e.g., MIT App Inventor's *Look*), but with other tools, users can construct or customize their own ML models (e.g., *Cognimates*, Google's *Teachable Machine*, Apple's *Create ML*, and *eCraft2Learn*). For example, students can use such tools to create custom image classification models that enable apps or games they are making to respond to objects in the world. These model construction tools typically emphasize the design of ML models via the definition of sets of classes (i.e., categories that define the vocabulary of models' predictions) and the creation of datasets that represent those classes, to be used for model training. However, the tools that typically support classification, but not regression or generation, have limited or no support for creating and using test sets or other mechanisms for systematic model evaluation, and offer little or no support for inspecting or modifying model architectures or hyperparameters. They also lack support for collaborative modeling activities, with Tseng et al.'s (2024, 2023) *Co-ML* being the rare exception (see Figure 3). Nor do most of these tools support making generative models, such as those used to create text, imagery, or other media based on natural language, an increasingly common way that people experience AI systems in everyday life (e.g., ChatGPT). These gaps are areas that future AI education tools should contend with.

The second, and far less common, category of tools are those that students can use to inspect and manipulate small-scale ML model architectures and, in the process, learn about neural networks and other technical approaches to AI. For example, Google's *Tensorflow*

*Playground* makes it possible for users to see neural network architectures so that they can experiment with how different combinations of input nodes, hidden layers, and output nodes affect an ML system's ability to learn different patterns in data. These tools are intended to support learning of specific technical ML concepts and are not designed for the kind of practical problem-solving tasks that can most motivate learning. Nonetheless, because they expose technical concepts that may be difficult for students to wrestle with in the context of applied problem solving, we believe that they represent an area of AI education tooling that merits further investigation.

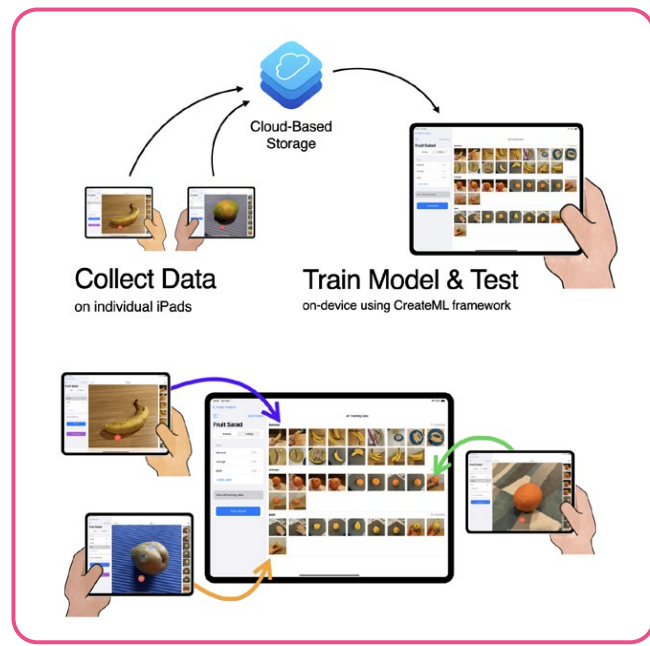

### Co-ML

Co-ML is a tablet-based app for beginners to collaboratively build image classifiers. Through a distributed experience that supports multiple devices, learners can iteratively refine ML datasets in discussion and coordination with their peers.

**Figure 3:** Co-ML: Collaborative Machine Learning Model Building (Tseng et al., 2023)

### [ML-Machine.org](ML-Machine.org)

ML-Machine is a web- and micro:bit-based educational tool designed to facilitate the broader integration of AI literacy in K-12 education. By enabling students to build and experiment with their own machine learning models, ML-Machine fosters computational empowerment—a capacity for students, both individually and in groups, to critically and creatively construct and deconstruct AI.

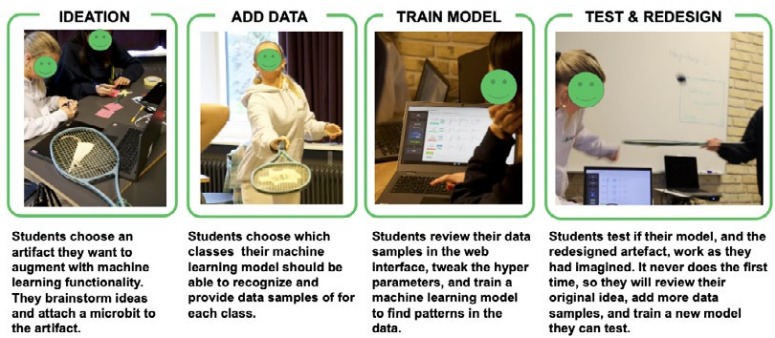

**Figure 4:** Children train their own ML model using ML-machine and Micro:bits

These tools (especially the first category) are used in educational settings in a variety of ways, including highly scripted, step-by-step pedagogies (e.g., Code.org's *AI Lab*); open-ended, creative, project-based learning (Fiebrink, 2019); and approaches that mix structured learning about AI/ML concepts with more applied exercises (e.g., Lee et al., 2021).

In the workshop meeting, we discussed the following questions:

- *What kind of tools already exist or need to be designed to support creative AI design activities?*
- *What are the design features that support students in different grade bands and with different motivations to work in Creative AI?*
- *How can we develop technologies that provide students with firsthand insights into machine learning, emphasizing the design choices made when designing technology?*
- *How can we design tools and experiences that help students move beyond traditional ideas of AI/robots?*
- *What are the existing benchmarks for measuring the quality of AI tools/models in education? How might we create policies concerning the auditing of new tools to pass validated benchmarks?*

In conceptualizing the innovation possibilities of tools for creating AI and using AI creatively, we considered ways to theorize, build, and study new tools, which we refer to as *CreateAI* tools. Discussions focused on how to support creative empowerment and learners' intrinsic motivation, how designs could build on students' knowledge and lived experiences, and how multimodal design could support situated learning on and off the screen and through tangible media. Questions were raised about the level of transparency for learners regarding the types of models (e.g., deterministic vs. nondeterministic) and layers of abstractions in a system. We also considered the role of teachers in the design, development, and use of these learning tools. Finally, community involvement was seen as central in design decisions at various stages (e.g., evaluation, ideation, decisions about data). Overall, we came to several conclusions, outlined below, about the kinds of tools that should be designed, who they should be designed for, and what factors should be considered in the design of these new tools.

### 2.1.1 Recommendations for CreateAI Tools

Participants made several recommendations about the kinds of tools that should be designed, who they should be designed for, and what factors should be considered in the design of these new tools.

When conceiving of new CreateAI tools, we need to **investigate how future tools address tradeoffs between malleability (or rigidity), robustness, and composability.** In other words, we need to better understand when AI learning tools should be more like Play-Doh or more like LEGO. While Play-Doh is infinitely malleable, it also is fragile, static, and hard to integrate with other materials. LEGO is stronger and is designed for building static and dynamic structures and mechanisms, but it also imposes hard constraints, as it's almost all right angles.

We also need to **support the development of tools that allow users to do progressively more complex investigations.** Many AI education tools are designed for beginners. They prioritize simplicity and ease of use. But when learners hit the limits of what they can do with these tools, there are no clear next steps for them to incrementally advance their skills. This lack of progressions has long been a problem in computing education: tools like Scratch and App Inventor allow users to create a multitude of projects, but when users wish to go beyond what these systems afford, there is no easy offramp toward more powerful tools. Researchers should investigate how to develop progressions of tools that avoid sharp cliffs (e.g., the massive drop off in productivity and growth in knowledge needed to transition between using App Inventor and using the raw Android SDK) and thus foster people moving from AI novices toward AI experts. Within this context it is also worthwhile investigating how to integrate AI in the design of applications. For instance, tools like [Replit](#) allow users to easily build their own websites that integrate AI/ML in a variety of ways.

Another focus should be on the users of CreateAI tools. We need to **envision CreateAI tools for everybody.** Just as most programmers today are not computer scientists—after all, the most used programming language in the world is Excel formulas—tomorrow's AI creators will come from all walks of life and create many different things for many different contexts. As we envision the kinds of progressions suggested above, we should not make the mistake of assuming that being an expert means being an ML engineer or another computer scientist. Creating tools for everybody also means developing tools for specific domains and disciplines that can be integrated into a rich set of cultures, practices, and communities. Novices could become expert artists or prostheticians who use AI to bring their visions to life. We also need to **help learners—including teachers—select appropriate tools**. Finding the right tool to create AI applications or be creative with AI is difficult. We need frameworks and resources to help people discover resources that are relevant to their goals, skill levels, and contexts.

As we are designing tools for creating AI or creating with AI, we need to consider several critical dimensions. First, CreateAI tool designers should **provide documentation of the types of projects their tools are intended to support** and the kinds of situations they have been tested to work well in. Ideally, they would also imagine and **disclose conditions and applications for which they have not been tested but for which the tools might foreseeably be used**. AI/ML systems often break down when applied in situations that they were not designed and tested for, such as when working with out-of-domain data. There is a risk that schools or other institutions will adopt AI tools only to find that they work poorly in particular contexts, for particular people, or for particular tasks. Here it is also worth noting that many AI tools are marketed as general purpose tools when, in fact, they are quite bad at particular tasks. For example, most LLM-based tools (e.g., chatbots) cannot do math well but may provide aesthetic outputs that falsely convey the impression that they do ([Solyst et al., 2024](#)). We need to support learners' and teachers' choices in using AI, but they should not expect that AI can do everything. Rather, learners and teachers need to learn to appreciate that some problems might require the use of calculators or other tools.

Second, we need to **disclose AI-based features**. AI methods are increasingly used to implement features within software applications that are not obviously "AI tools." This can lead to surprising or even problematic results (e.g., [Samsung phones that add false details to pictures of the sky](#)). Tools developed for schools should disclose what functionality is AI based and how that functionality was created and evaluated. Researchers should investigate how stakeholders, including school/district leaders, teachers, parents, and students, learn to reason about the potential implications—including possible errors—of these implementation decisions.

Finally, we need to **support linguistic and other cultural heterogeneity**. Just as educators must learn to value the vernacular practices that different students bring to the classroom, future CreateAI tools should be designed and tested to work well for students who speak, write, draw, or otherwise communicate in ways that differ from dominant norms. This includes robust attention to accessibility so that people who speak, see, hear, or otherwise sense and express in atypical ways may be just as creatively empowered as those with more typical abilities. Otherwise, new AI tools for learning may perpetuate old structures of domination and oppression.

Overall, we need to carefully examine and design for desirable social futures with AI. As noted above, few CreateAI tools today support collaborative modeling (Co-ML is the sole novice-focused example that we are aware of). Yet much of human productivity, and all of human learning, is social. Tools research should **take into account the multiscale social dynamics of CreateAI futures**, from social practices of modeling or model use to the dynamics of classrooms, libraries, and homes where those CreateAI activities happen. This includes careful attention to power dynamics within those spaces, and problematic incentives and punishments they can impose. Some companies are already adopting performance metrics that incentivize greater reliance on AI and penalize less use of it. We should **resist creating circumstances where a student who intentionally chooses not to use AI in a project is judged to be performing worse than students who do.** As policymakers increasingly prioritize AI literacy and mastery, we should develop assessment methods that do not remove student or teacher agency to resist AI use when they find it inappropriate for their goals.

# 2.2. LEARNING

We already have a substantial body of research examining students' comprehension of AI/ML systems. The majority of these studies concentrate on how young people assess AI/ML in terms of their intelligence and human-like traits such as adherence to rules and trustworthiness (e.g., Druga et al., 2017, 2018, 2019; Williams et al., 2019). Many of these attributions are influenced by popular media portrayals. While these studies offer valuable insights into students' perceptions of ML technologies, most of them do not address how learners develop an understanding of the internal mechanisms of these technologies or how learners acquire a conceptual grasp of the ML pipeline (Fiebrink, 2019)--in other words, an understanding the relationships between training data, learning algorithms, and models. Recent studies by Druga and Ko (2021) and Szczuka et al. (2022) have shed more light on children's comprehension of ML and explored their misconceptions. They also investigated how novices tend to overestimate the capabilities of AI/ML technologies due to inaccurate or simplified mental models (e.g., Hitron, 2019). Additionally, scholarly investigations have explored how students can construct models (for instance, Adams et al., 2023; Alvarez et al., 2022) and cultivate proficiencies in artificial intelligence and machine learning (DiPaola et al., 2022; Long & Magerko, 2021). Yet developing these roadmaps of students' conceptions is particularly challenging in the current landscape because what's available constantly changes. For instance, the computing systems themselves are changing, and this affects the tools and what parts of the computing systems learners and teachers have access to and how. Students' conceptions could be rooted in particular, sometimes poor, design decisions of the tools/systems. And as systems become more complex and opaque, understanding what role AI/ML has within the computing system might also be muddled.

During our workshop conversations, we focused on the following questions:

- *What do we need to know about students' prior understanding at different age levels? Is anthropomorphizing in AI/ML/LLM a hindrance or a blessing? What learning designs can effectively demystify AI and help students develop scientifically grounded models of AI systems?*
- *How do we empower students and teachers to feel comfortable using and creating AI in a manner that enhances creativity rather than stifling it?*
- *Can we enable students to invent or remodel AI systems? What can students learn about a subject domain and AI systems by building a domain-specific computational creativity system?*
- *How do we make sure youth have a voice in this process and in answering the preceding questions?*

We engaged in robust, multidisciplinary conversations about the potential and pitfalls of using generative AI in learning contexts. Discussions emphasized the importance of designing AI tools that support—rather than replace or supplant—student agency, creativity, and critical thinking. Participants shared examples, such as DataBites (Figure 5), that provided hands-on activities in a museum setting to help youth interrogate data bias and reflect on AI system design. Workshop participants identified core challenges such as preventing over-dependence on AI tools, fostering intentional "friction" to deepen learning, and resisting the narrative of AI as an inevitable, monolithic force. Ethical concerns around data quantification, surveillance, and the amplification of systemic biases were also recurrent themes, with a call for more culturally responsive AI learning experiences.

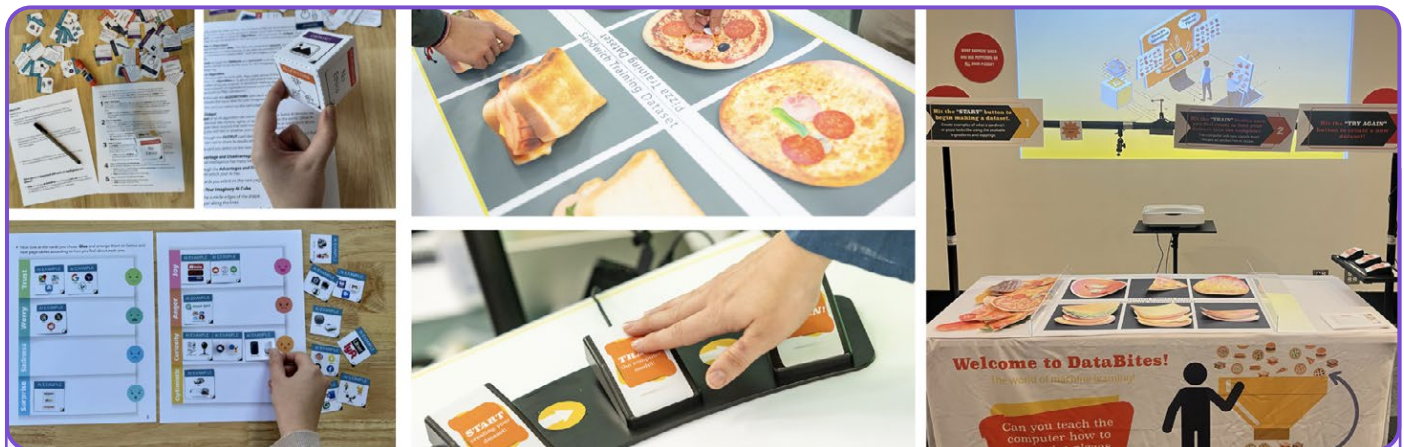

Darabipourshiraz and colleagues (2025) designed AI literacy experiences to engage young learners in understanding AI and its ethical implications. Through museum exhibits (such as DataBites), unplugged activities, and interactive web applications, they provide diverse, accessible entry points into AI education. These experiences promote active exploration and creation, encouraging students to move beyond passive use and critically examine AI systems. By connecting AI to ethics, creativity, and real-world relevance, the work deepens understanding of how AI functions and affects society. It equips students with the tools to question, shape, and contribute to AI technologies in thoughtful and informed ways.

**Figure 5:** (Left) Three images of an unplugged activity. (Right) Three images of the DataBites museum exhibit

### 2.2.1 Recommendations for CreateAI Learning

As a result of workshop discussions, participants recommended that researchers and developers of CreateAI tools and activities **focus on what prior funds of knowledge to build on and what scaffolds to develop so that learners can develop a better understanding of different aspects**, such as the design of data sets or training of models; this is especially important given the lack of transparency in system operations. We also need to understand what creative work students want to do. Questions about agency should be addressed in regard to ownership of learning, especially when learners co-design with AI/ML technologies.

Beyond focusing on students' comprehension of concepts, **we also need a better grasp of the learning that students engage with when creating AI/ML applications**. For instance, most research in this area focuses on how youths train models, with much less attention to the equally important task of testing models when creating machine learning applications. Understanding the practices students engage with when building data sets, training and testing models, and incorporating them into applications is key to developing curricula. Learners should not just engage in the training but also the evaluation of AI/ML systems as part of their creative process.

Furthermore, we need to **engage youth in learning about ethics and criticality in ways that are authentic, empowering, and relevant throughout the design process.** For this we need to develop a better understanding of how young people make sense of various AI technologies from ethical and technical perspectives. Ultimately, the group advocated for a vision of AI education where students become co-creators and critics of AI systems; where they not only learn with AI but also about it—including its design, limitations, and implications for justice and relations of power.

# 2.3. ETHICS

The majority of research on understanding ethics in AI/ML applications has considered rights and safeguards (Blikstein & Blikstein, 2021; Ito et al., 2021; Santer et al., 2021), high-stakes policing and surveillance technologies (e.g., Vakil & McKinney de Royston, 2022; Walker et al., 2022), and hypothetical interactions with robots (e.g., Charisi et al., 2021). Only a handful of studies have delved into young people's views of algorithmic justice in their everyday experiences (Schaper et al., 2022). For instance, Coenraad (2022) and Salac and colleagues (2023) explored how young people perceive the intersections of technology, ethics, and algorithmic fairness, while Salac and collaborators (2023) presented scenarios of algorithmic unfairness to children (ages 8-12) to stimulate their understanding of system operations. Ruppert (2023) investigated youths' perceptions of surveillance of them by governmental and commercial actors in schools, neighborhoods, and online spaces, with Ruppert and colleagues (2023) describing how playful workshop structures support youth learning to actively resist such surveillance.

During our workshop, conversations moved beyond surface-level understanding of AI ethics and into the nuanced practices of data, fairness, and accountability.

Here are some of the questions we took up:

- *How can we ensure fairness and justice in algorithmic systems? What mechanisms can be developed to identify and mitigate harmful bias in AI?*
- *How can educational approaches help K-12 students understand the ethical implications of AI technologies? What experiences and pedagogical methods can foster critical thinking about AI's societal impacts?*
- *What are the fundamental limitations of AI technologies? How can we educate students to critically assess and shape the broader impacts of AI on society and the environment?*
- *How can AI education balance technological creativity with ethical responsibility? How can we create engaging, hands-on learning experiences that explore AI ethics without over-relying on technological devices? What collaborative and inquiry-based methods can help students deeply understand AI's ethical challenges?*

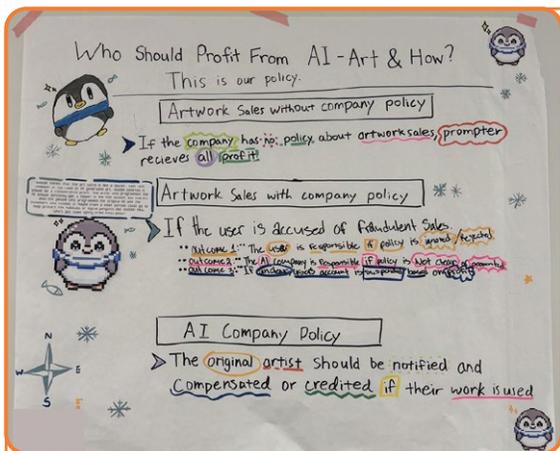

**Empowering Youth Agency in AI Design: Policy Design to Foster Critical AI Literacy in K-12 Students**

Ali and colleagues (2025) engaged 94 middle and high school art students in a policy-design learning activity as a part of an Art and AI learning workshop. Students worked in groups to create policies around AI's use in art, considering stakeholders like artists, AI companies, and consumers. Students developed nuanced, actionable policies that reflected a deep understanding of AI's impact on the art ecosystem, including issues of copyright, artist compensation, and transparency. The activity empowered students to think critically about AI's ethical implications on the AI and art ecosystem and fostered a sense of citizenship and agency in shaping AI futures. This work demonstrates the value of integrating policy design activities into AI curricula, providing youth with the skills and perspectives to become informed, ethical citizens in an AI-driven future. functional understandings related to how the quality and quantity of the training data and the training process influence the quality of the outputs. They considered ethical issues related to the trustworthiness of model outputs, the appropriate use of language models, authorship, and the ownership of creative ideas.

**Figure 6:** *Policy Design Activity:* Student poster demonstrating their beliefs on attribution of AI-generated artwork.

In one example presented by attendees (Figure 1), researchers guided teenagers to build and reflect on their own mini language models ("babyGPTs") as a way for them to grapple with how data sources and training decisions impact model outputs. Here youth engaged in iterative design processes that involved technical development as well as collaborative critique regarding ethical dilemmas that are embedded in generative AI. In another presentation (Morales-Navarro et al., 2024), researchers described how youth explored their role as auditors of AI systems, learning how to critically evaluate the outputs and internal logic of platforms like YouTube, TikTok, and ChatGPT. Overall, we note that less attention has been paid to how youth can examine algorithmic bias in the design of their own machine learning applications, although some recent work has started to integrate the perspectives of different stakeholders (see Ali et al., 2021) (see Figure 6) or examined peer testing and auditing of wearable machine learning applications (Morales-Navarro et al., 2024). By engaging in two strands of inquiry—design and evaluation—young people illustrated how they can meaningfully shape how AI is developed. Through these examples, our workshop highlighted the power of playful and participatory approaches to foster youth agency in understanding and designing algorithmic futures. Finally, we note that ethics are rarely addressed in current computing curricula. While numerous educational programs have been created to assist educators in introducing AI/ML subjects to their students (for instance, Druga et al., 2022), only a small subset of these programs tackle issues related to social and ethical considerations. This observation also became evident when examining reviews of K-12 machine learning curricula (Rauber & Gresse von Wangenheim, 2023; Sanusi et al., 2023).

**AI Bias Impeding Real-Time Co-Creative Ideation for Black Students' Conceptual Game Design**

Edouard (2024) examined how racial and cultural biases in AI image generators impede the creative process for Black students engaged in conceptual design. In this case study, Devonte, a Black undergraduate game design student, used AI to bring his hand-drawn characters to life. Despite clear visual and textual prompts that included descriptors like "Black male with glowing blue hair," the AI consistently misrepresented his intent by lightening skin tones, straightening culturally significant hairstyles, and altering eye color. Devonte was forced to iterate repeatedly, often re-entering terms like "Black man" multiple times, to partially achieve his intended design. This friction not only disrupted his workflow but revealed how AI systems resist Black cultural representation. Edouard's work highlights the need for AI literacy that includes critical interrogation of algorithmic bias and supports students' cultural agency in design spaces. It demonstrates the importance of empowering marginalized youth to critique and reimagine AI tools that often exclude them.

Top: Initial AI output that misrepresented Devonte's character by altering skin tone and hairstyle.
Bottom: Text-prompted version created after multiple attempts to reflect his intended cultural aesthetic.

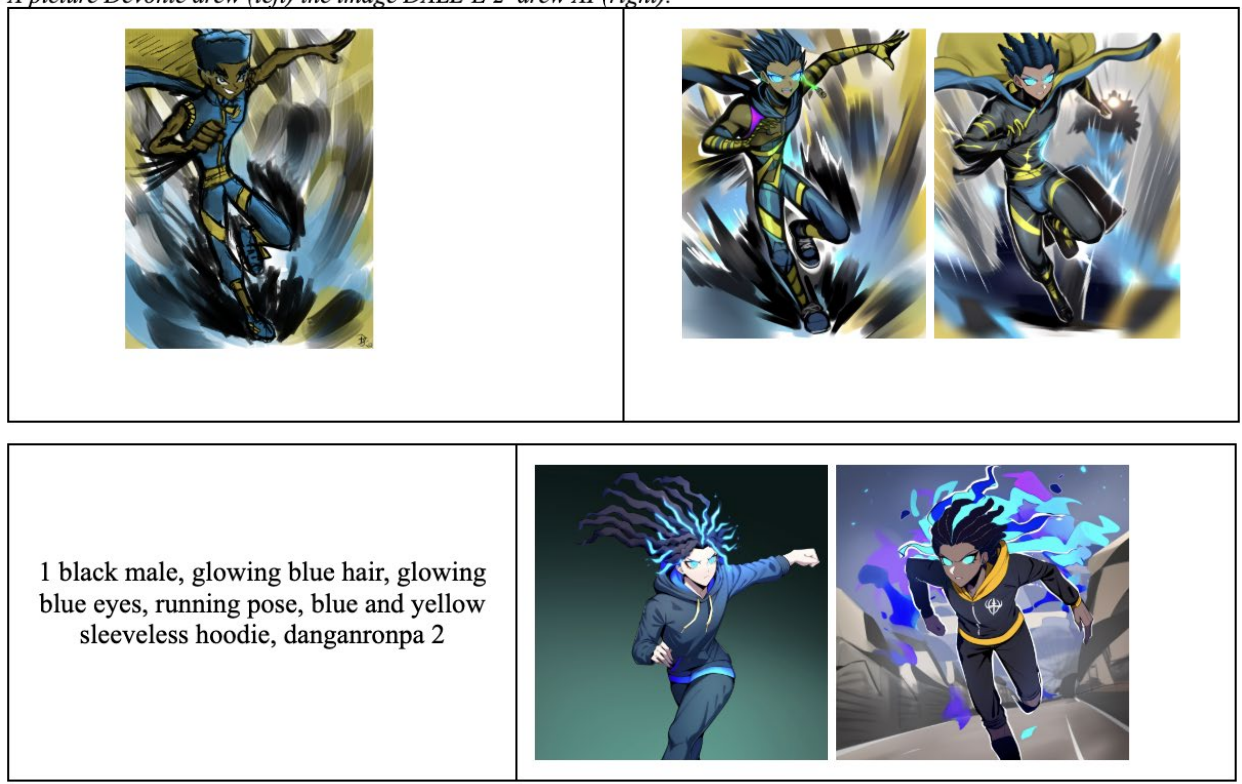

**Figure 1**
*A picture Devonte drew (left) the image DALL-E 2 drew AI (right).*

**Figure 7:** Examples of Real-Time Co-Creative Ideation in Game Design with AI

### 2.3.1 Recommendations for CreateAI Ethics

As a result of workshop discussion, participants recommended that ethics should (1) be integrated into course content, (2) provide relevant experiences, and (3) be taught using a variety of approaches. First and foremost, there was consensus among all participants about the need to **integrate ethics throughout curricular activities rather than treating them as an add-on or afterthought**. Understanding the limitations of AI/ML systems should be considered an opportunity to examine AI's functionally and interrogate the design of datasets and algorithms, as well as the intentions of the designers of systems. Here we need to help students understand the origins of data, how models are used and created, what values they represent, and how responsibility is distributed across the landscape. Research also should **investigate which experiences can help students to grapple with the limitations of AI and where those limitations come from** so they can critically assess what shapes the societal, environmental, and ethical impacts of AI/ML. AI/ML models (including the models created by learners!) are a representation of a set of values and norms through which AI/ML systems reproduce/create/transform society. How are designers considering these implications in their work? Finally, there is a need to **create and investigate pedagogical approaches and activities that are hands-on (i.e., not relying solely on discussion or lectures), collaborative and inquiry-based** and can help students understand AI's ethical impact and limitations. Here, activities such as algorithm auditing can provide a productive approach for students without prior coding experience to examine AI/ML systems. We also need to investigate further how students can be involved in critiques and audits and develop ways for them to share their audit results.

# 2.4. TEACHING

Conspicuously little attention has been paid to teachers' knowledge and strategies in K-12 AI/ML education. A handful of investigations have documented how teachers view and execute AI-related tasks (e.g., Chounta et al., 2022; Sanusi et al., 2023; Velander et al., 2023; Williams et al., 2021), offering preliminary glimpses into their grasp of teaching ML concepts and methodologies, as well as the hurdles faced by students. Studies have also documented that teachers have limited understandings of algorithmic bias, unfairness, and technological solutionism (Hu & Yadav, 2023; Zhang et al., 2022). For instance, Lin and Van Brummelen (2021) underscored the specific challenges that teachers encounter when incorporating ethical subjects into their instruction concerning AI/ML.

We must also build on what computer science (CS) teachers already know and teach in current K-12 computing classrooms. Considerable efforts have focused on inducting new teachers into K-12 computing, but this work typically does not include AI/ML content and ideas. Any effort to bring AI/ML topics into K-12 CS education needs to build on what teachers understand about computational thinking and what has recently been encoded in K-12 CS standards. In the United States, the Computer Science Teachers Association (CSTA) is currently determining (see Figure 8) how and what AI content should be included in foundational CS standards (CSTA & AI4K12, 2025), while internationally the UNESCO (UNESCO, 2023) and OECD/EC (OECD/EC, 2025) frameworks provide guidance. Figuring out what can be done within existing standards, or with only minimal modifications to those standards, is likely to matter a lot.

In the workshop, we addressed questions such as:

- *What are the most promising pedagogical approaches for introducing CreateAI concepts in K-12 schools and informal learning environments? How can existing computing education frameworks be meaningfully adapted to incorporate AI and machine learning concepts?*
- *What age-appropriate and accessible AI-learning activities can be developed for K-12 students? How will computational thinking (CT2.0; see also Tedre et al., 2021) need to transform to address data-driven problem-solving in an AI-integrated world? What new mental models are crucial for students to develop in understanding AI systems? What strategies can help students understand the practical implications of AI technologies?*
- *What approaches can schools use to leverage generative AI as an empowerment tool in low-income educational spaces? How can AI education minimize the risk of perpetuating systemic inequities while fostering student creativity?*
- *How can educators with limited technical backgrounds be prepared to teach AI concepts effectively?*

Koressel shared a progress update on the Computer Science Teachers Association's (CSTA's) work to update its K-12 Computer Science Standards. The CSTA K-12 Standards are intended to define what students should know and be able to do in computer science by the end of each grade level/band. They serve as the de facto national standards in the United States, and most states either adopt them or heavily base their own standards on them. The updated standards will be informed by both research and practice, form coherent progressions that can be implemented flexibly based on contextual factors and needs, and represent a future-oriented view of a comprehensive K-12 CS education. The updated CSTA K-12 Standards will be released in summer 2026.

**Figure 8:** Timeline for the CSTA K-12 Standards revision process.

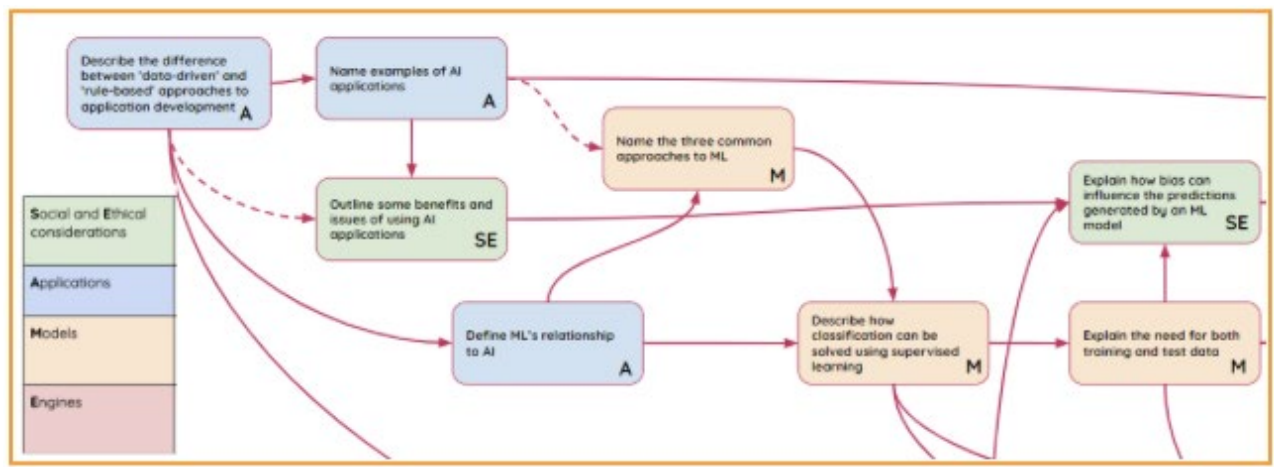

Waite and colleagues (2024, 2025) presented about Experience AI, a collection of resources for learning about and teaching AI that were developed with Google DeepMind. As well as lesson plans and teaching materials, teachers are provided with learning graphs that give a learning progression. A set of design principles underpins the resources, including developing a core vocabulary, avoiding anthropomorphism, using the SEAME framework to examine various levels of development (i.e., Social and Ethical considerations, Application, Model, Engine), focusing on data-driven rather than rule-based problem solving (computational thinking 2.0), and threading real-world examples throughout. Materials have been localized for delivery in dozens of countries worldwide, including translation work and changing activity contexts.

**Figure 9:** Part of the learning graph for an Experience AI lesson, the SEAME framework, and recommendations for avoiding the use of anthropomorphizing language when describing AI systems.

The workshop highlighted the urgent need to support educators in navigating, designing, and critiquing AI-integrated learning environments through a multidimensional lens. Key examples showcased research-informed teacher learning and curriculum resources (Figure 9) that help educators introduce foundational AI concepts, address issues of ethics and equity, and design unplugged activities (see Figure 10).

Other examples illustrate how AI activities can become part of the humanities classroom (see Figure 11). Participants discussed how teachers function not only as implementers but as co-designers, advocates, and learners, especially as they confront a landscape where AI is simultaneously presented as a tool for creative expansion and as a potential substitute for instructors.

### AI as a Creative Tool

Romeike (see Issa et al., 2025) presented a number of approaches for bringing AI to K-12 CS classrooms as a creative tool — for example, by developing a computational creativity system (CCS). CCSs are systems that produce a creative product in a creative domain: By designing and creating a simplified unplugged or semi-plugged CCS, students engage their own creativity to produce creative artifacts with AI techniques, helping them understand how AI can be "creative."

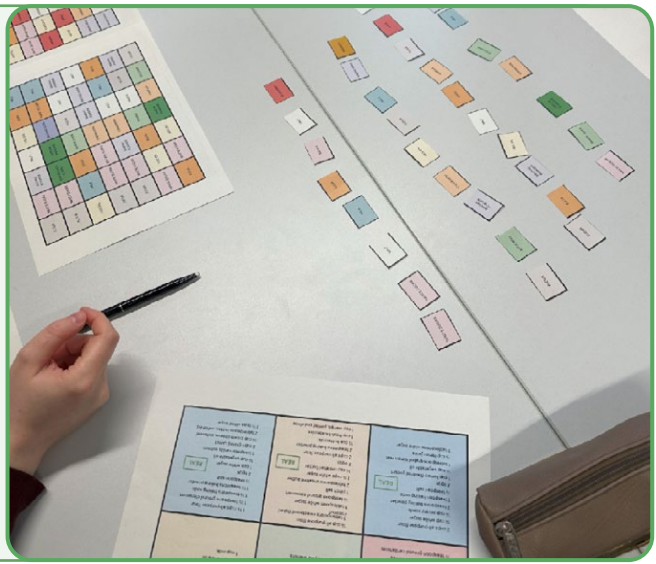

**Figure 10:** Building computational creativity systems.

Nichols and colleagues (see Thrall et al., 2024) presented early findings from a study that partnered with teachers in an urban public high school to explore how humanities classrooms can be sites for considering how emerging technologies are intermediating civic life. Pairing research on platform ecologies (Nichols & Garcia, 2022) and civic inquiry, they shared a provisional framework for digital civic inquiry that guided curriculum and teaching in the study and discussed implementation challenges and opportunities.

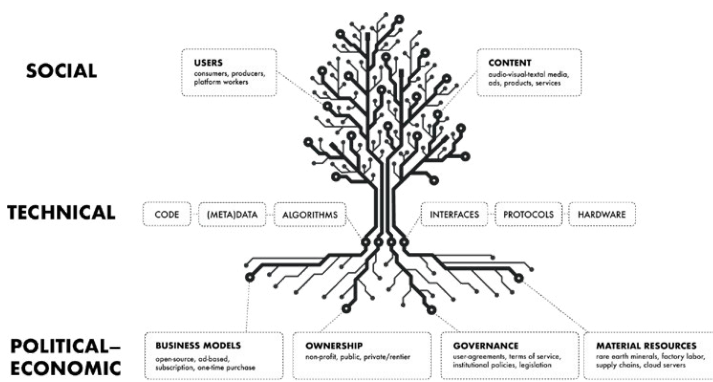

**Figure 11:** (Top) Visualization of a platform ecology; (Bottom) Provisional digital civic inquiry framework, co-designed by researchers and teachers.

### 2.4.1 Recommendations for CreateAI Teaching

The discussions underscored that to make the teaching of CreateAI approaches in K-12 classrooms a reality, multiple challenges must be addressed. Workshop participants recommended that a major focus should be on providing teachers with professional development, materials, and tools, but further research is also needed to address varying district policies around AI/ML applications. To begin, this requires **research into which pedagogical approaches are most effective for teaching CreateAI to students of different ages and backgrounds.** We also need to develop a better understanding of whether teachers should use data-driven rather than rule-based programming, and whether to use no-code rather than coding approaches to promote knowledge development. Furthermore, we need to **identify the forms of professional development and resources teachers need to facilitate student learning experiences centered on creating AI/ML applications.** Here, teacher professional development also needs to be expanded, acknowledging the steep learning curve associated with AI's technical complexity and constant changes. Ethics and sociotechnical critique also need to be embedded throughout professional development to help make visible failures, frictions, and biases inherent in AI systems to foster critical reflection. Finally, given that the current policy fragmentation across districts and agencies remains an obstacle, we must **provide schools and teachers with clarity about which tools can be purchased and accessed, where AI/ML can be integrated (given that curricular time is limited), and guidance on what works well.**

# 2.5. ASSESSMENT

Assessment of AI/ML applications designed by learners is an underdeveloped area. Most studies use interviews that probe students' thinking processes and reflections during hands-on projects and their understanding of AI concepts. But this approach is resource intensive and feasible with only a small number of students (Tang et al., 2020), making it an unrealistic tool for classroom applications. Researchers have also evaluated student portfolios of AI projects and programs. For example, Kaspersen and colleagues (2021) evaluated students' AI models and user interface designs. After analyzing the artifacts in students' projects, the researchers found that children were able to apply their new knowledge of machine learning to their own lives and to brainstorm personally meaningful applications using ML. While such portfolios are informative, there is a need for grading rubrics that capture dimensions of AI learning performance. Surveys with quantitative items and open-ended questions can also be used to gauge students' perceptions of AI, confidence, and readiness. For instance, Register and Ko (2020) applied qualitative thematic analysis of students' open-ended responses about how machine learning systems work, as well as other aspects including ML model transparency, critical thinking, and learners' interests and backgrounds.

Overall, our discussions about assessing learners' CreateAI applications focused on the following questions:

- *How can teachers assess student learning of AI key concepts and practices in creating AI applications? What comprehensive assessment methods can educators develop to evaluate students' understanding of AI's core conceptual foundations? How can assessment go beyond technical proficiency to measure critical thinking about AI's broader societal and environmental implications?*
- *What rubrics and assessment tools can effectively measure students' ability to create meaningful AI applications? How can educators design assessment strategies that capture both the technical skills and the critical reflection required in AI development?*
- *How can educators critically examine the ecological consequences of AI technologies alongside their innovative potential?*

In line with the workshop theme of *Creating with AI*, researchers have also designed new tools that use AI to help students document their own projects. One such example is the Make-In-Progress tool, which students can use to document their efforts and then create AI-assisted presentations of their projects (Danzig et al., 2025b) (see Figure 12). In a related example (Danzig, Correa, & Holbert, 2025a) developed an analytical framework consisting of three categories of elements—object, scene, and context—that multimodal AI makes inferences about. They then tested the framework in a series of mini-experiments using ChatGPT 4.0 with GPT-4v, using work generated by students engaged in extended making experiences. While interviews and portfolios can provide valuable insights into students' conceptual understanding and practices and are appropriate for early stage research, they are insufficient tools for gauging students' knowledge. Overall, the development of validated knowledge tests is far behind where we need it to be.

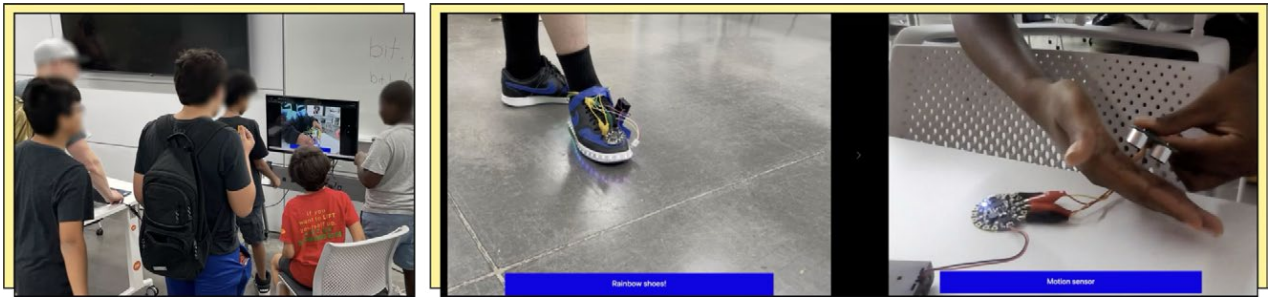

With Make-in-Progress (Danzig et al., 2025b), students can quickly and easily capture photos or videos of their work, add a brief caption, and instantly publish them to carousel displays located in geographically distributed makerspaces. Additional multimodal generative AI features support makers and educators in probing the database. For example, students might ask the system to generate a slideshow of their extended maker project, or a teacher might ask the system to identify new materials that a student group might need for their project.

**Figure 12:** Images taken by students and displayed on the Make-in-Progress photo carousel.

Solyst et al. (2025) explored how young people could detect biases and other harmful issues in AI through auditing different types of systems. They found that youth could detect four different types of problematic AI outputs: misinformation, harmful bias, technical limitations, and inappropriate content. This work contributes to our understanding of how youth can actively engage in learning about human–AI alignment by exploring the technical and socio-technical limitations of AI. The team prototyped an auditing tool that youth could use to compare and contrast different text-to-image generative AI outputs to learn and audit AI models.

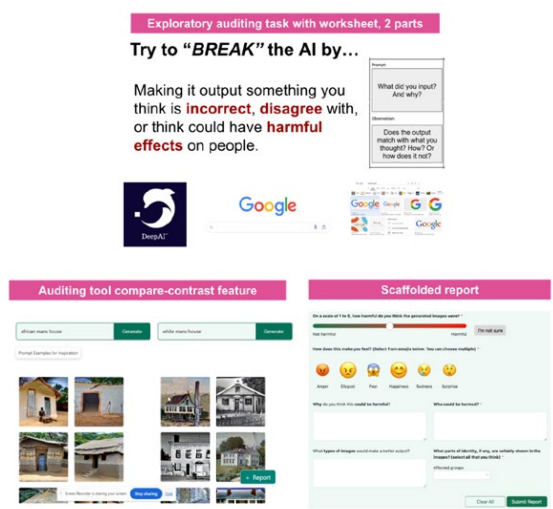

**Figure 13:** The auditing task (top) is to "break"—or explore the limitations of—three different types of AI models. A screenshot of the auditing tool's compare-contrast interface (bottom left) and scaffolded report system (bottom right) that supports reflection and surfacing of a problematic AI model behavior.

### 2.5.1 Recommendations for CreateAI Assessment

Workshop participants emphasized the need to (1) establish baselines, (2) design different instruments, and (3) develop approaches that promote students as responsible creators of AI. First, we need to **establish baselines of core AI/ML concepts that students need to know in order to create AI applications.** We also need a deeper understanding of this work across different grade bands and contexts. Second, we need to **investigate different approaches and instruments for assessment.** One approach could be to develop student-based self-assessments. For instance, surveys could help capture what students have learned at the end of the day and be both feed-back and feed-forward mechanisms (using questions such as What did I learn? What will I need to focus on?). In the context of portfolio-based assessments, students could select the best sample of what they have achieved and explain why. These assessments could also focus on possible benefits that students derive from completing a project. Another approach could be to focus on artifact-based assessments. For instance, students could conduct peer audits (a form of algorithm auditing; see Morales-Navarro et al., 2024), focusing on functionality and ethics, as a way to give feedback. Students could also apply an ethical stakeholder matrix to assess how their applications look from the perspective of a company, advertiser, parents, another student, and other stakeholders. Finally, workshop participants expressed a strong interest in **developing approaches that promote students as responsible creators of AI.** They wanted to find ways for learners to interrogate the impact of their own AI developments but wondered how much technical content knowledge students would need to engage critically.

# 3. NEXT STEPS FOR CREATEAI

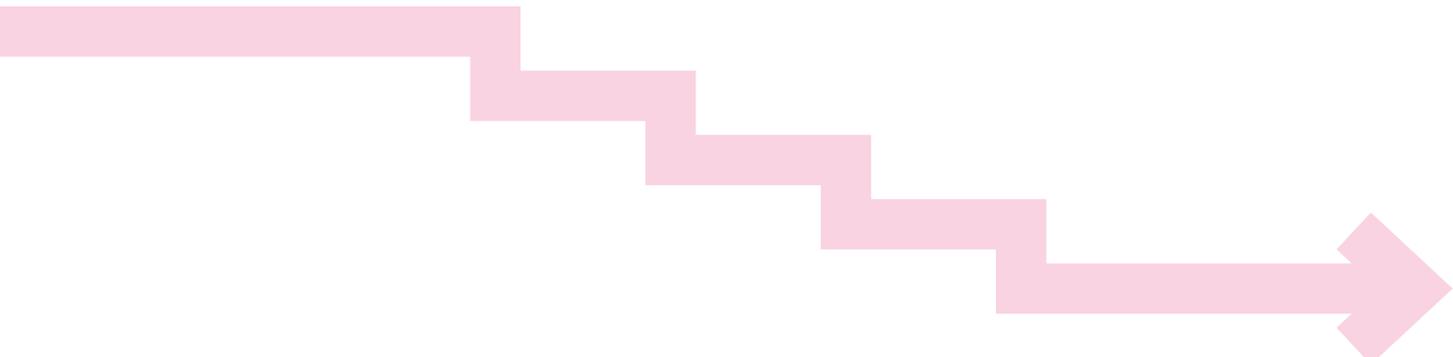

In this workshop, AI and computing education experts and practitioners explored the opportunities and challenges of preparing students for a world in which AI/ML applications are part of their 21st-century education (see Figure 12). Our discussions revealed that fostering creative agency and critical thinking requires deliberate pedagogical approaches that demystify AI systems. By framing these technologies as designed artifacts with inherent limitations and biases—rather than as objective or infallible tools—learners can develop a more nuanced understanding of how design choices embed values and limitations into technologies that shape human experiences and social structures (see also Figure 14). As standards in computing education are updated in light of recent AI/ML developments, we hope that the ideas discussed in this workshop will help shape the recommendations.

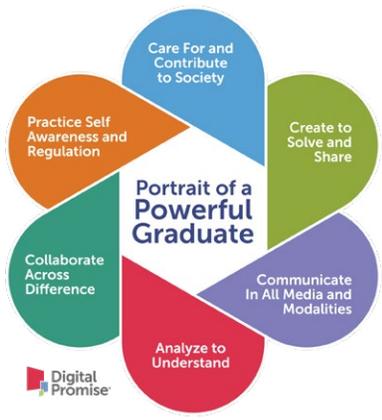

**Portrait of a Powerful Graduate** establishes a vision for learners' future success. Many school districts and states have developed such portraits to articulate qualities learners should exemplify by the time they leave formal schooling.

Digital Promise is exploring how emerging technology can promote Portrait of a Graduate competencies such as creativity and collaboration through curriculum, pedagogy, and assessment.

**Figure 14:** Portrait of a Graduate with AI literacies

To advance policy and practice, we highlight key features of AI/ML education that reflect the aims of CreateAI:

**Provide Teaching Support.**

The rapid expansion of AI tools has left teachers grappling with implementation demands that exceed available support structures. Today, educators must navigate complex technological landscapes while meeting pedagogical requirements—all without sufficient scaffolding. Any new development, curricula, or tool should be accompanied with robust support for teachers to learn about AI themselves and teach about AI to their students.

**Promote Accessibility.**

Current approaches frequently overlook the diverse learning needs. Effective AI integration requires intentional design for students with learning disabilities, visual impairments, and other accessibility requirements. Without such a comprehensive approach, AI tools risk becoming barriers rather than empowering resources.

**Address Ethical Responsibilities.**

The psychosocial dynamics between AI systems and young users warrant heightened scrutiny. Designer responsibility must address the emotional dimensions of human–machine interaction. As children and adolescents increasingly form emotional attachments to conversational agents and digital assistants, the ethical implications of fostering such connections demand critical examination. These relationships, while seemingly benign, may fundamentally reshape social development patterns and interpersonal expectations in ways not fully anticipated by system designers.

**Develop Challenge-Centered Pedagogies.**

Several participants advocated for approaches where students are encouraged to either develop intentionally faulty AI systems or systematically attempt to compromise existing ones. This approach represents a departure from traditional educational paradigms that emphasize building increasingly robust systems. By engaging with AI systems at their points of failure, students develop a nuanced understanding of the technological limitations that might otherwise remain abstract or theoretical. When students witness firsthand how an AI system fails—whether through intentional design or through systematic probing—they develop an embodied understanding of the technology's constraints. As students deliberately induce system failures, they confront the potential real-world implications of such failures in deployed contexts.

**Realize Environmental Impact.**

The rapid proliferation of artificial intelligence systems requires the exploration of environmental consequences. Despite growing awareness of the resource-intensive nature of large AI models, the ecological footprint of AI development and deployment remains insufficiently quantified. This absence of standardized metrics impedes meaningful comparative analyses and obscures the differential environmental burden imposed by AI infrastructure across diverse socioeconomic landscapes. It is key for learners to engage with, discuss, and consider the environmental implications of developing and using AI.

In promoting *CreateAI* we advocate moving beyond knowing how to use AI systems to developing a multidimensional framework in which meaningful AI literacy includes being able to create AI and create with AI. We want to promote not only technical understanding but also political and personal dimensions of engagement with AI technologies. The personal dimensions address how learners cultivate their voices and agency in relation to these technologies. This aspect of literacy enables individuals to leverage AI as a means of self-expression and personal fulfillment rather than experiencing it as an external force that diminishes human creativity or autonomy. The equally important political dimension incorporates an understanding of how AI technologies reshape civic participation and collective decision-making processes.

# APPENDIX

## *APPENDIX OVERVIEW*

## *APPENDIX A: VIRTUAL MEETUPS IN FALL 2024*

### October 4, 2024 Learning and Teaching with Creating AI

Panelists: Netta Livari (University of Oulu), Kayla DesPortes (New York University), & Thomas Philip (UC Berkeley)
Moderators: Kathi Fisler (Brown University) & Jayne Everson (University of Washington)

### October 18, 2024 Designing Tools for Creating AI

Panelists: Tiffany Tseng (Barnard College), Jody Medich (Google), & Vishesh Kumar (Vanderbilt University)
Moderators: Ben Shapiro (University of Washington) & Luis Morales-Navarro (University of Pennsylvania)

### December 14, 2024: Envisioning Learning Futures with Creating AI

Panelists: Arturo Cortez (University of California, Berkeley) & Tiera Tanksley (UCLA)
Moderator: Jose Ramón Lizárraga (University of California, Berkeley)

## APPENDIX B: MARCH 2025 WORKSHOP ATTENDEES

Marina Bers, *Boston College*
Francisco Castro, *New York University*
Hasti Darabipourshiraz, *Northwestern University*
Kayla DesPortes, New York University
Stefania Druga, Google

Kareem Edouard, *Drexel University*
Jayne Everson, *University of Washington*
Deborah Fields, *Utah State University*
Ole Sejer Iversen, *Aarhus University*
Allen Sejer Iversen, *Thorning Skole*

Nathan Holbert, *Columbia University*
Sarah Judd, *Franklin School*
Yasmin Kafai, *University of Pennsylvania*
Jake Koressel, *CSTA*
Anssi Lin, *Eastern Finland University*

Netta Iivari, *University of Oulu*
José Ramón Lizárraga, *UC Berkeley*
Mike Mead, *Apple*
Kelly Mills, *Digital Promise*
Luis Morales-Navarro, *University of Pennsylvania*

Phil Nichols, *Baylor University*
Daniel Noh, *University of Pennsylvania*
Thomas Philip, *UC Berkeley*
Stuart Ralston, *Apple*
Ralf Romeike, *Freie Universität Berlin*

Carolyn Rose, *Carnegie Mellon University*
Ricarose Roque, *University of Colorado Boulder*
R. Benjamin Shapiro, *University of Washington*
Jaemarie Solyst, *University of Washington*
Matti Tedre, *Eastern Finland University*

Tiffany Tseng, *Barnard College*
Jane Waite, *Raspberry PI Foundation*
Marcelo Worsley, *Northwestern University*

# *APPENDIX C: WORKSHOP AGENDA MARCH 2-4, 2025*

**SUNDAY, MARCH 2**

5:00 PM - 5:40 PM Poster Session I

Carolyn Rosé, Jayne Everson, Francisco Castro, Jane Waite, T. Philip Nichols, Kelly Mills, Ole Sejer Iversen, Thomas M. Philip, Luis Morales-Navarro, Ricarose Roque, Tiffany Tseng, Marcelo Worsley

5:45 PM - 6:15 PM Why this workshop?
Panel: Yasmin Kafai, R. Benjamin Shapiro, Marina Bers, José Ramón Lizárraga

6:20 PM - 7:00 PM Poster Session II
Kareem Edouard, Deborah A. Fields, Hasti Darabipourshiraz, Kayla DesPortes, Jaemarie Solyst, Nathan Holbert, Jake Koressel, Netta Iivari, Daniel J. Noh, Ralf Romeike, Matti Tedre/Anssi Lin

**MONDAY, MARCH 3**

8:30 AM - 9:00 AM Welcome & Overview
9:00 AM - 9:30 AM Fireside Chat I: What is CreateAI?
Panel: Stefania Druga, Matti Tedre, Kayla DesPortes

9:30 AM - 10:45 AM Working Group I: Defining "CreateAI"

11:00 AM - 12:00 PM Working Group Reconvene I
Reporting back on "CreateAI" & Literacy

1:30 PM - 3:30 PM Working Group II: Topics of CreateAI
Tools; Learning; Teaching; Assessments

3:30 PM - 4:00 PM Working Group Reconvene II
Tools; Learning; Teaching; Assessments

4:30 PM - 5:30 PM Fireside Chat II: AI in Europe
Panelists: Netta Iivari (Finland), Ole Sejer Iversen (Denmark), Ralf Romeike (Germany), & Jane Waite (UK)

**TUESDAY, MARCH 4**

8:30 AM - 9:45 AM Demo Party!
Panelists: Stefania Druga, Matti Tedre, & Marcelo Worsley

9:50 AM - 10:45 AM Working Group III: Topics of CreateAI
11:00 AM - 12:15 PM Working Group Reconvene III
1:15 PM - 2:15 PM Group Reconvene: Ethics
2:15 PM - 3:00 PM Open Questions & Conclusions

# *APPENDIX D: ISLS 2025 Conference Symposium*

*Youth as Designers of Artificial Intelligence and Machine Learning Technologies: What Do We Know About the Opportunities and Challenges of K-12 Students Creating Their Own Applications?*

**SYMPOSIUM CHAIRS**
Yasmin B. Kafai, University of Pennsylvania,
R. Benjamin Shapiro, University of Washington

**SYMPOSIUM PRESENTATIONS**
*Making AI-Based Apps Strengthens Children's Data Agency and Understanding of AI*
Matti Tedre, Henriikka Vartiainen, University of Eastern Finland

*Empowering Students to Shape the Future: Understanding AI/ML Learning Difficulties in K-12 Education*
Franz Jetzinger, Tilman Michaeli, Technical University of Munich

*Critical, Ethical, Empowering Design of AI/ML Systems for Social Good*
Netta Ilvari, University of Oulu

*Empowering Teachers as Users and Creators of Specialized AI Tools for Diverse Classrooms*
Safinah Ali, New York University

*Understanding and Creating with AI: Tools and Strategies for Computing Education*
Line Have Musaeus, Ole Sejer Iversen, Aarhus University

*Re-Situating the Salience of Context in Designing for Authentic Engagement with AI/ML-Based Design for Athletics and Play*
Herminio Bodon, Meg Butler, Khushbu Kshirsagar, Vishesh Kumar, Michael Smith, Ashley Quiterio, Marcelo Worsley, Northwestern University

*Building your Own Generative Language Model: High School Students' Engagement with Data Practices and Ethical Considerations when Designing babyGPTs*
Luis Morales-Navarro, Daniel J. Noh, & Yasmin B. Kafai, University of Pennsylvania

**SYMPOSIUM DISCUSSANTS**
Roy Pea, Stanford University
Thomas Philip, University of California Berkeley